\documentclass[12pt]{article}
\usepackage{graphicx}
\renewcommand{\baselinestretch}{1.5}

\begin{document}

{ \baselineskip = 2pc

\begin{center}
{\bf EFFECTIVE SCREENED POTENTIALS OF STRONGLY COUPLED
SEMICLASSICAL PLASMA}
\end{center}
\begin{center}
{\bf T.S. Ramazanov, K.N. Dzhumagulova\\}
   {\it  Al Farabi Kazakh National University, IETP, \\
     Tole bi, 96, 480012, Almaty, Kazakhstan}
\end{center}

\begin{verse}
                 Corresponding author: Prof. Dr. T.S.Ramazanov\\
                 Tole bi, 96a,\\ 480042, Almaty, Kazakhstan \\
                        Fax: 7(3272) 50-39-79\\
                        E-mail: ramazan@physics.kz;\\
                                dzhkn@physics.kz
\end{verse}

{\bf Abstract. } The pseudopotentials of particle interaction of a
strongly coupled semiclassical plasma, taking into account both
quantum-mechanical effects of diffraction at short distances and
also screening field effects at large distances are obtained. The
limiting cases of potentials are considered.


 {\bf I. Introduction.}

At the present time the study of strongly coupled plasmas
properties is of great interest, it caused by fundamental
investigations of astrophysical objects  and inertial confinement
fusion plasmas \cite{dewitt}-\cite{ichi}. In such strongly coupled
systems the collective (screening) and quantum-mechanical effects
play an important role in studies of thermodynamic and kinetic
properties of strongly coupled plasmas. Furthermore, in kinetic
theory which is based on kinetic equations, there are problems
with divergence of collision integrals at big and small scattering
angles. The one way to eliminate them is the using of the bare
Coulomb potential, being cutted at large and short distances. On
the other hand, it is known that the divergence at small angles
appears as a result of not taking into account the plasma's
screening effects on interparticles scattering in the system. The
other method of eliminating divergence at long ranges and
correctly describing the properties of strongly coupled plasmas is
to use the effective potentials of interaction of charged
particles. These potentials take into account the influence of
surrounding particles, it leads to the screening of external
potential $\varphi(r)$. Therefore, the strongly screened effective
potential that takes into account three-particle correlations is
obtained in \cite{nur}. Notice that this potential does not take
into account quantum mechanical effects and describes the
interaction between particles of classical dense plasma only.

In this paper the effective potential for semiclassical plasma is
obtained. This potential contains quantum diffraction effects at
short distances as well as screening effects for large distances.

    {\bf II. The system parameters.}

The fully ionized plasma consisting of electrons and ions is
considered. Number density is considered in the range of
$n=n_e+n_i=10^{20}-10^{24}cm^{-3} $, and the temperature domain is
$10^{5}-10^7 K$. The average distance between particles (or the
Wigner-Seitz radius) is : $$a=(\frac{3}{4\pi n})^{1/3},$$ where
$n_i=Zn_e$ are number densities of ions and electrons. Other
parameter characterizing the state of system is coupling
parameter:
 $$\Gamma=(Ze)^2/(ak_BT),$$
 where $T$ is plasma temperature, $k_B$ denotes the Boltzmann
 constant,
  $\Gamma$ is the ratio between average Coulomb interaction energy and thermal energy.
For strongly coupled plasma $\Gamma>1$.

Density parameter is defined as
 $$r_s=a/a_B,$$
 where $a_B=\hbar^2/(m_ee^2)$ is Bohr radius, $r_s$ increases with decreasing of density.

The degree of Fermi degeneracy for the electrons is measured by
the ratio: $$ \Theta
=\frac{k_BT}{E_F}=2(\frac{4}{9\pi})^{2/3}Z^{5/3}\frac{r_s}{\Gamma},
$$ where $E_F$ is the Fermi energy of electrons. The condition
$\Theta\geq 1$ corresponds to the state of weakly and intermediate
degeneracy.

    {\bf III. Pseudopotentials.}

It is well -known that the electrical field of particles in plasma
induces the density fluctuations of charges, i.e. the polarization
effect is evolved. As a consequence, the opposite charged "cloud"
around test charged particle is formed, it leads to screening of
initial potential \cite{noz}. The relationship between the total
(screened) potential and externally applied one is as follows:

\begin{equation} \label{Phi}
\Phi(q)=\frac{\varphi(q)}{\varepsilon(q)},
\end{equation}

where  $\Phi(q),\varphi(q)$ are Fourier transforms of the screened
and external potentials, respectively. Fourier transformation is:
\begin{equation} \label{fi}
\varphi(q)=\frac{4\pi}{q}\int_0^\infty r\varphi(r)sin(qr)dr,
\end{equation}
$\varepsilon(q)$ is the dielectric function of static screening.
In the random-phase approach for weakly degenerate plasma
$\varepsilon(q)$ can be derived from following relation \cite{ih}:
\begin{equation} \label{eps}
\varepsilon(q)=1+\sum_{\alpha}\frac{4\pi n_\alpha
\varphi(q)}{k_BT}
\end{equation}

In case of the Coulomb potential, we have:
\begin{equation} \label{fic}
\varphi_c(q)=\frac{4\pi e^2}{q^2}
\end{equation}
\begin{equation} \label{epsc}
\varepsilon_c(q)=\frac{q^2+\kappa^2}{q^2},
\end{equation}
where $\kappa^2=(4\pi ne^2)/(k_BT)\equiv 1/r_D^2$ is the value
that is inverse proportional to the square of  Debye radius. Using
Eqs. (\ref{Phi}),(\ref{fic}),(\ref{epsc}),  Fourier transform of
effective potential $\Phi(q)$ can be obtained.  The inverse
Fourier transformation
\begin{equation} \label{Phir}
\Phi (r)=\frac{1}{2\pi ^2r}\int_0^\infty q\Phi (q)sin(qr)dq,
\end{equation}
gives well-known Debye-H\"uckel potential:
\begin{equation} \label{DH}
\Phi_{DH}(r)=\frac{e^2}{r}e^{-\kappa r} .
\end{equation}
Hence the divergence at the large distances is avoided by taking
into account screening effects, but it still remains at short
distances. However, it is known that taking into consideration the
quantum-mechanical effects leads to the finite values of
pseudopotential at short distance. This method is called the
method of thermodynamic pseudopotential, and the effective
temperature dependent pseudopotential is obtained as a result of
correlation between the classical Boltzmann factor and the
quantum-mechanical Slater sum \cite{dun}. In this framework  the
potential taking into consideration effects of diffraction has
been derived for high densities and temperatures \cite {kel},
\cite{de},\cite{ux}
\begin{equation} \label{kdy}
\varphi_{\alpha \beta}(r)=\frac{Z_{\alpha}Z_{
\beta}e^2}{r}(1-e^{-r/\lambda_{\alpha \beta }})
\end{equation}
where $\lambda=\hbar/\sqrt{2\pi \mu_{\alpha \beta }k_BT}$ is the
thermal de Broglie wavelength of $\alpha -\beta $ pair,
$\mu_{\alpha \beta }$ is the reduced mass of  $\alpha - \beta $
pair. Potential (\ref{kdy}) is not screened at large distances. It
raises the idea to obtain the effective potential on the basis of
the dielectric responce function as a result of screening the
potential (\ref{kdy}). Applying
Eqs.(\ref{fi}),(\ref{eps}),(\ref{Phi}) to (\ref{kdy})  we have
obtained:
\begin{equation} \label{epsq}
\varepsilon (q)=\frac{q^4+q^2/\lambda_{\alpha
\beta } ^2+\kappa^2/\lambda_{\alpha \beta
}^2}{q^2(q^2+1/\lambda_{\alpha \beta } ^2)},
\end{equation}
\begin{equation} \label{Phiq}
 \Phi (q)=\frac{4\pi
Z_{\alpha}Z_{\beta}e^2(q^2+1/\lambda_{\alpha \beta }
^2)}{q^4+q^2/\lambda_{\alpha \beta } ^2+\kappa^2/\lambda_{\alpha
\beta }^2},
\end{equation}
The pseudopotential $\Phi_{\alpha \beta }(r)$ can be restored from
(\ref{Phir}),(\ref{Phiq}) :
\begin{equation} \label{dd}
\Phi_{\alpha \beta }(r)=\frac{
Z_{\alpha}Z_{\beta}e^2}{\sqrt{1-4\lambda_{\alpha \beta
}^2\kappa^2}}\lambda_{\alpha \beta }^2(\frac{e^{-Ar}}{r}-
\frac{e^{-Br}}{r})
\end{equation}
where $$B^2=\frac{1+\sqrt{1-4\lambda_{\alpha \beta
}^2\kappa^2}}{2\lambda_{\alpha \beta }^2},$$
$$A^2=\frac{1-\sqrt{1-4\lambda_{\alpha \beta
}^2\kappa^2}}{2\lambda_{\alpha \beta }^2}$$

It is obvious that pseudopotential (\ref{dd}) is valid when
$4\lambda_{\alpha \beta }^2\kappa^2<1 $ (or $\lambda_{\alpha \beta
}<r_D/2$), i.e. in the region of weakly degenerate plasmas.

Since potential (\ref{kdy})  behaves as the Coulomb potential at
large distances it is natural to suppose that dielectric function
is defined by Coulomb's relation (\ref{epsc}). In this approach,
using expression (\ref{epsc}) for $\varepsilon (q)$ instead of
(\ref{epsq}), one can obtain the expression for effective
potential:
\begin{equation} \label{dc}
\Phi_{\alpha \beta }(r)=\frac{
Z_{\alpha}Z_{\beta}e^2}{1-\lambda_{\alpha \beta
}^2\kappa^2}(\frac{e^{-\kappa r}}{r}-\frac{e^{-r/\lambda_{\alpha
\beta }}}{r})
\end{equation}

Let us consider the limiting cases of the expressions
(\ref{dd}),(\ref{dc}).

Case 1.  $\kappa\rightarrow 0 (r_D\rightarrow \infty).$  From the
physics point of view it means that there are no screening
effects. Then potentials (\ref{dd}) and (\ref{dc}) coincide with
(\ref{kdy}).

Case 2. In the absence of diffraction effects
($\lambda_{\alpha\beta}\rightarrow 0$), from (\ref{dd}) and
(\ref{dc}) we have the potential (\ref{DH}).

Case 3. When both screening and diffraction effects are absent
($\lambda_{\alpha\beta}\rightarrow 0, \kappa\rightarrow 0 $),
pseudopotentials (\ref{dd}) and (\ref{dc}) coincide with the
Coulomb potential.

Case 4. $\lambda_{\alpha\beta} \kappa << 1
(\lambda_{\alpha\beta}<<r_D).$ In this case pseudopotential
(\ref{dd}) coincides with potential (\ref{kdy}) and limiting
formula of pseudopotential (\ref{dc}) is as follows:
\begin{equation} \label{dclim}
\Phi_{\alpha \beta }(r)=\frac{
Z_{\alpha}Z_{\beta}e^2}{r}(e^{-\kappa r}-e^{-r/\lambda_{\alpha
\beta }}) .
\end{equation}
Similar limiting expression have been derived in \cite{dav} on the
basis of the linear chain of Bogolyubov's equations in  pair
correlations approach by the micropotential  taking into
consideration diffraction and symmetry effects. It should be noted
here that results of \cite{dav} are valid for weak coupled plasmas
( $\Gamma <1$).

The  pseudopotentials (\ref{dd}) and (\ref{dc}) as a function of a
distance between particles for the different values of parameters
$r_s$ and $\Gamma$ are presented on the figures. It is shown that
both potentials  coincide with the Debye-H\"uckel potential at
$r\rightarrow \infty$  and   they have finite value at
$r\rightarrow 0$.
   Discrepancy between  (\ref{dd}) and (\ref{dc}) grows with increasing
of coupling parameter at fixed $r_s$ (Figs.1 , 2 and 3 ) and also
with increasing of density at fixed $\Gamma$ (Figs.3,4). Under the
same conditions the discrepancy between limiting expression (13)
and pseudopotentials (11) , (12) grows too (Figs.5,6 and 7).

Consequently, pseudopotentials  derived in this work take into
account both quantum-mechanical effects of diffraction at short
distances and also screening field effects at large distances
between particles. Last time some authors  have calculated
potentials which take into consideration quantum-mechanical
effects and collective events \cite{dav},\cite{bek1},\cite{bek2}
by different methods. However, analytical expression for
pseudopotential which is valid in wide region of $r_s$ and $\Gamma
$ has not been yet obtained.

In conclusion it should be noted here that the expressions for
pseudopotentials derived in this work are simple enough and can be
easily used in analytical calculations and computer simulations in
investigations of strongly coupled semiclassical plasma
properties.

\newpage

\large
\renewcommand{\baselinestretch}{1.0}

\newpage
\vspace*{3cm}
\bigskip
\centerline{\bf  Figure captions }
\bigskip
\vspace{1cm}

\noindent Figure 1:
 Effective potentials of the interaction between particles of a
 semiclassical hydrogen plasma. Solid line: formula (\ref{dc});
 circles denote the formula (\ref{dd}); dashed line: potential (\ref{kdy}); dashed-dotted line
 represents the Debye-H\"uckel potential.

\vspace{2mm}

\noindent Figure 2:
 The same as on Fig.1.

\vspace{2mm}

\noindent Figure 3:
The same as on Fig.1.

\vspace{2mm}

\noindent Figure 4:
The same as on Fig.1.

\vspace{2mm}

\noindent Figure 5:
Effective potentials of the
interaction between particles of a
 semiclassical hydrogen plasma. Solid line: formula (\ref{dc});
 circles denote the formula (\ref{dd}); triangles
 represent the potential (\ref{dclim}).

\vspace{2mm}

\noindent Figure 6:
The same as on Fig.5.

\vspace{2mm}

\noindent Figure 7:
The same as on Fig.5.

\newpage

\vspace{2cm}

\begin{figure}[h]
\includegraphics[width=15cm,height=15cm]{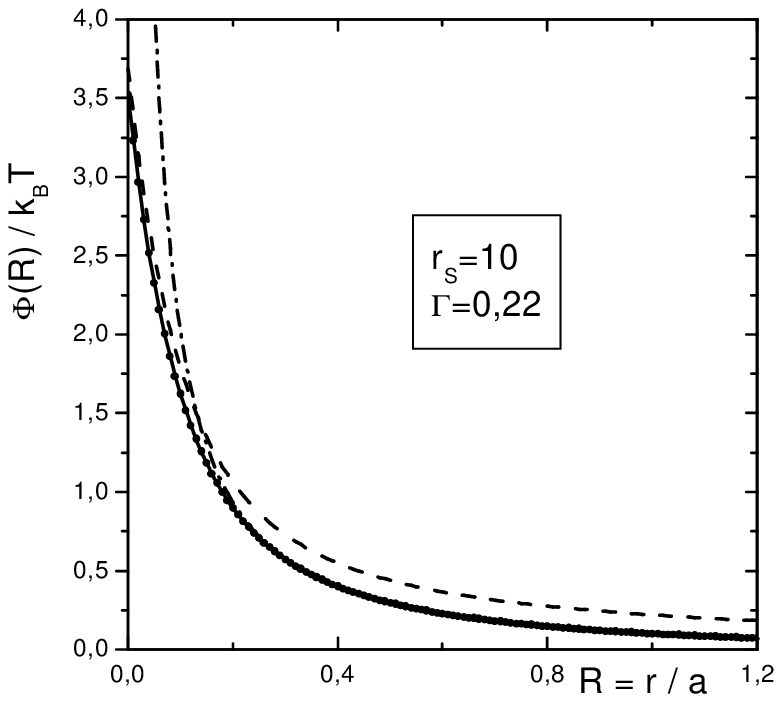}
\end{figure}
\vspace{2cm} T.S.Ramazanov, K.N.Dzhumagulova \\
\noindent Figure 1: {\small Effective potentials of the
interaction between particles of a
 semiclassical hydrogen plasma. Solid line: formula (\ref{dc});
 circles denote the formula (\ref{dd}); dashed line: potential (\ref{kdy}); dashed-dotted line
 represents the Debye-H\"uckel potential.}

\newpage

\vspace*{2cm}

\begin{figure}[h]
\includegraphics[width=15cm,height=15cm]{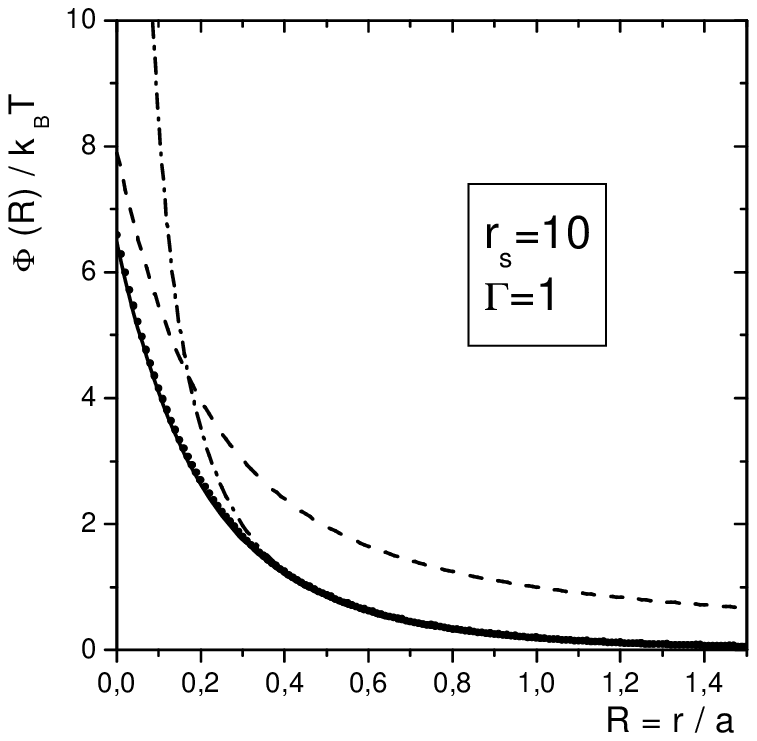}
\end{figure}
\vspace{2cm} T.S.Ramazanov, K.N.Dzhumagulova \\
\noindent Figure 2: {\small The same as on Fig.1.}

\newpage

\vspace*{2cm}

\begin{figure}[h]
\includegraphics[width=15cm,height=15cm]{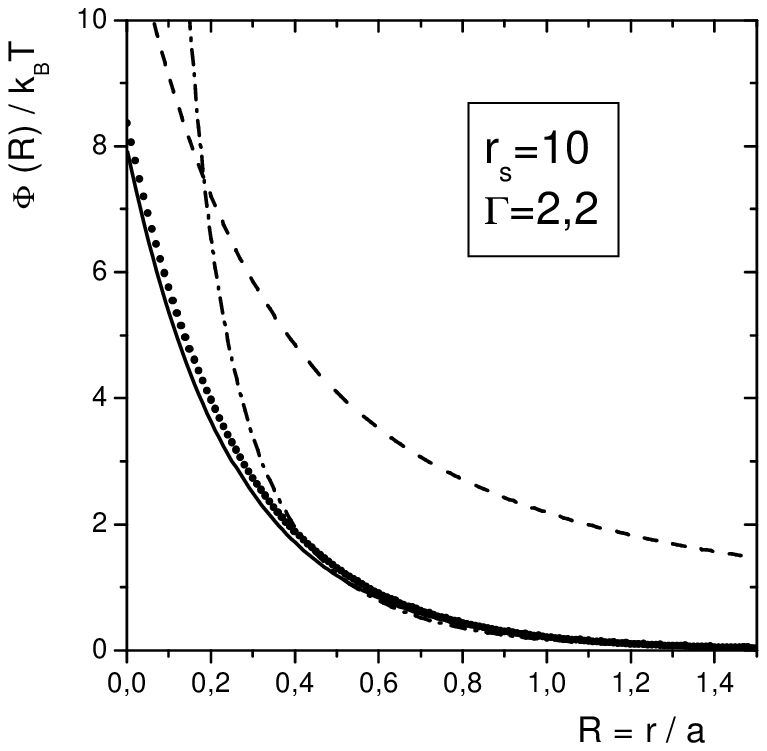}
\end{figure}
\vspace{2cm} T.S.Ramazanov, K.N.Dzhumagulova \\
\noindent Figure 3: {\small The same as on Fig.1.}

\newpage

\vspace*{2cm}

\begin{figure}[h]
\includegraphics[width=15cm,height=15cm]{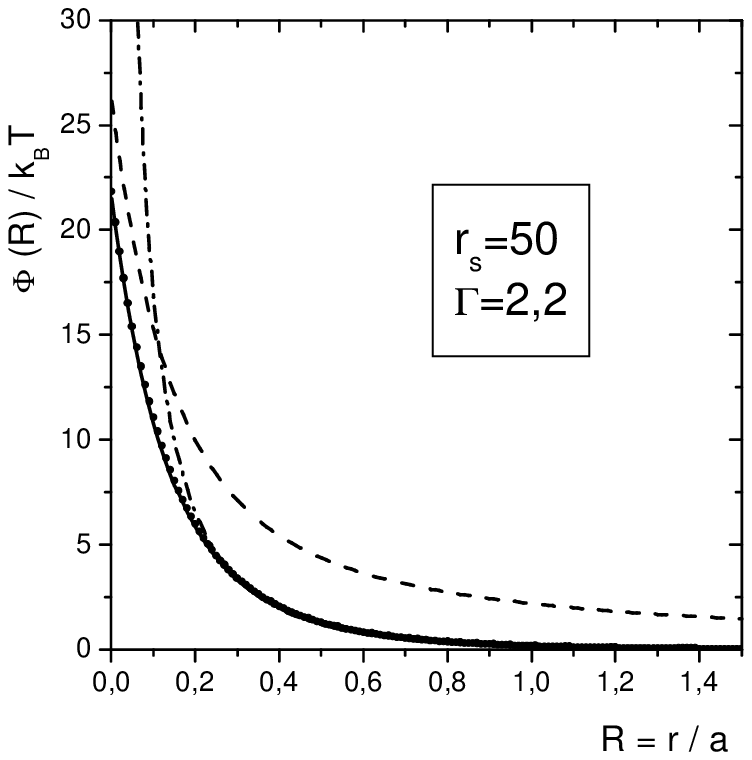}
\end{figure}
\vspace{2cm} T.S.Ramazanov, K.N.Dzhumagulova \\
\noindent Figure 4: {\small The same as on Fig.1.}

\newpage

\vspace*{1.5cm}

\begin{figure}[h]
\includegraphics[width=15cm,height=15cm]{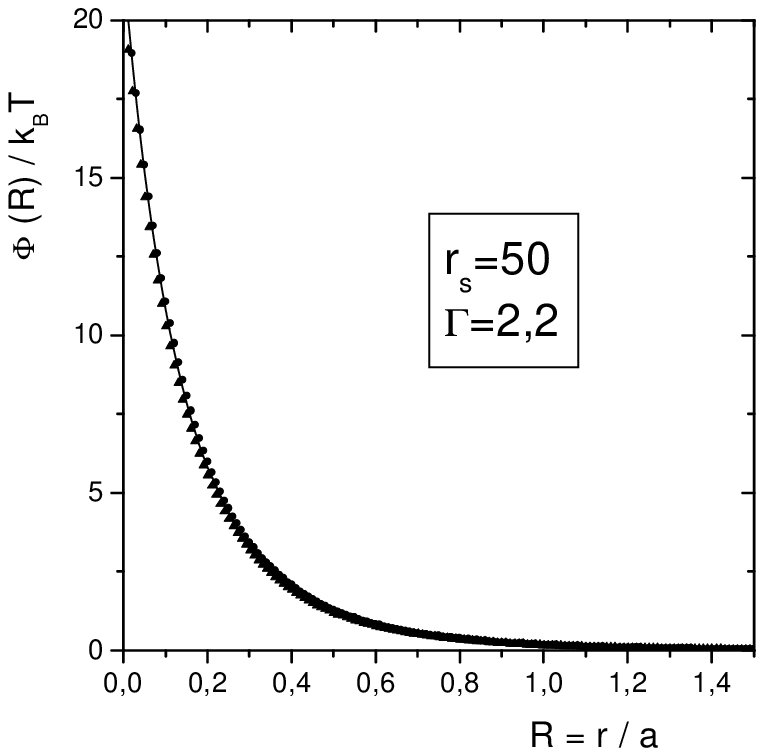}
\end{figure}
\vspace{2cm} T.S.Ramazanov, K.N.Dzhumagulova \\
\noindent Figure 5: {\small Effective potentials of the
interaction between particles of a
 semiclassical hydrogen plasma. Solid line: formula (\ref{dc});
 circles denote the formula (\ref{dd}); triangles
 represent the potential (\ref{dclim}).}

\newpage

\vspace*{2cm}

\begin{figure}[h]
\includegraphics[width=15cm,height=15cm]{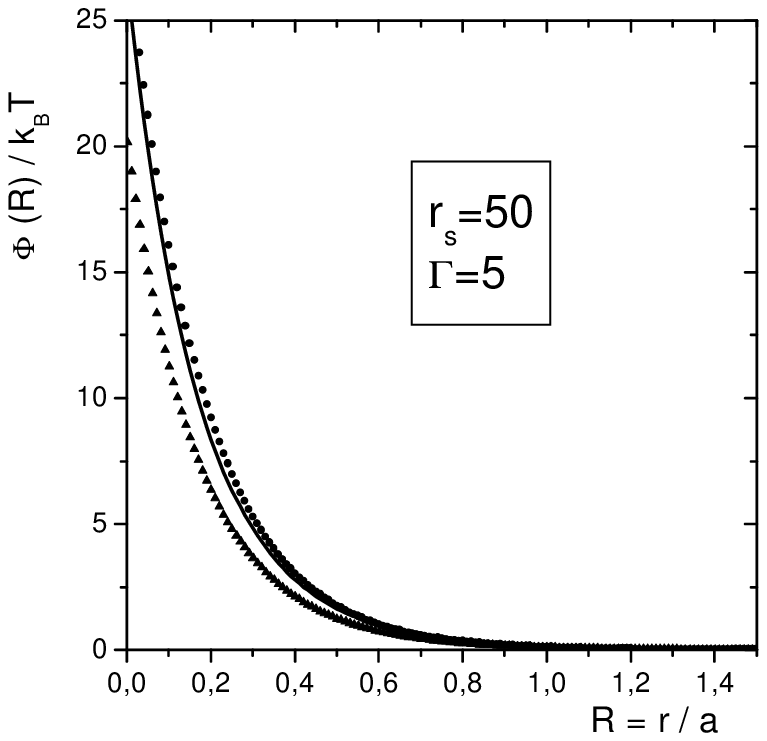}
\end{figure}
\vspace{2cm} T.S.Ramazanov, K.N.Dzhumagulova \\
\noindent Figure 6: {\small The same as on Fig.5.}

\newpage

\vspace*{2cm}

\begin{figure}[h]
\includegraphics[width=15cm,height=15cm]{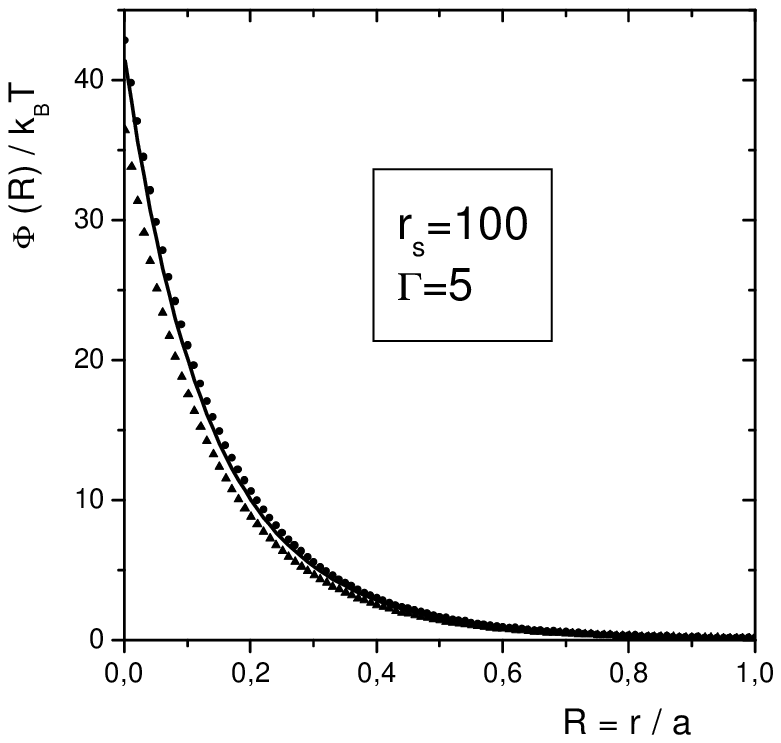}
\end{figure}
\vspace{2cm} T.S.Ramazanov, K.N.Dzhumagulova \\
\noindent Figure 7: {\small The same as on Fig.5.}

\end{document}